\documentclass[12pt]{article}
\usepackage{latexsym}
\usepackage{amsbsy}

\begin{document}

\def\WF{{\rm WF}}
\def\bfeta{{\mbox{\boldmath{$\eta$}}}}
\def\R{{\cal R}}
\def\I{{\rm i}}

\title{Quantum Field Theory on Spacetimes with a Compactly Generated 
Cauchy Horizon}  

\author{Bernard S.~Kay (bsk2@york.ac.uk) \\
Marek J.~Radzikowski$^*$ (radzikow@mail.desy.de) \\
\small{Department of Mathematics, University of York, Heslington, York
YO1 5DD, U.K.} \\
\small{($^*$Present address: II. Institut f\"ur Theoretische Physik,
Universit\"at Hamburg,} \\
\small{Luruper Chaussee 149, D-22761 Hamburg, Germany)} \\
\\
Robert M.~Wald (rmwa@midway.uchicago.edu) \\ 
\small{Enrico Fermi Institute, University of Chicago,  Chicago, Illinois
60637, U.S.A.}}

\date{Received: 14 March 1996/ Accepted: 11 June 1996} 

\maketitle

\begin{abstract}
We prove two theorems which concern difficulties in the formulation of
the quantum theory of a linear scalar field on a spacetime,
$(M,g_{ab})$, with a compactly generated Cauchy horizon.  These theorems
demonstrate the breakdown of the  theory at certain {\it base points} of
the Cauchy horizon, which are defined as `past terminal  accumulation
points' of the horizon generators. Thus, the theorems may be interpreted
as giving support to Hawking's `Chronology Protection Conjecture',
according to which the laws of physics prevent one from manufacturing a
`time machine'. Specifically, we prove:

\vspace{.5em}\noindent
{\bf Theorem 1.} {\it There is no extension to $(M,g_{ab})$ of the usual
field algebra on the initial globally hyperbolic region which satisfies
the condition of F-locality at any base point.  In other words, any
extension of the field algebra must,  in any globally hyperbolic
neighbourhood of any base point, differ from the algebra one would
define on that neighbourhood according to the rules for globally
hyperbolic spacetimes.}

\vspace{.5em}\noindent
{\bf Theorem 2.} {\it The two-point distribution for any Hadamard state 
defined on the initial globally hyperbolic region must (when extended to
a distributional bisolution of the covariant Klein-Gordon equation on
the full spacetime) be singular  at every base point $x$ in the sense
that the difference between this two point distribution and a local 
Hadamard distribution cannot be given by a  bounded function in any
neighbourhood  (in $M \times M$) of $(x,x)$.}  
\vspace{.5em}

In consequence of Theorem 2, quantities such as the renormalized
expectation value of $\phi^2$ or of the stress-energy tensor are 
necessarily ill-defined or singular at any base point.

The proof of these theorems relies on the `Propagation of Singularities'
theorems of Duistermaat and H\"ormander. 
\end{abstract}

\section{Introduction} \label{intro}

In recent years, there has been considerable interest in the question
whether it is possible, in principle, to manufacture  a `time machine'
-- i.e., whether, by performing operations in a bounded region of an
initially `ordinary' spacetime, it is possible to bring about a
`future' in which there will be closed timelike curves.  Heuristic
arguments by Morris, Thorne and  Yurtsever \cite{MorThorYur} suggested
in  1988 that this might be possible with a suitable configuration of
relatively moving wormholes,  and alternative ideas also have been
suggested by others (see e.g., \cite{FroNov}).  For reviews and further
references, see e.g., \cite{ThorneCord,Visbk}.

A precise, general class of classical spacetimes, $(M,g_{ab})$,   in
which a time machine is `manufactured' in a  nonsingular manner within
a bounded region of space is comprised by those  with a {\it compactly
generated Cauchy horizon} \cite{Haw92}.  By this is meant, first, that
$(M,g_{ab})$ is time orientable and possesses a closed achronal
edgeless set $S$ (often referred to as a partial Cauchy surface) such
that $D^+(S) \neq I^+(S)$, where $D^+(S)$  denotes the future domain of
dependence of $S$, and $I^+(S)$ denotes the  chronological future of
$S$.  (Often we shall refer to  the full domain of dependence, $D(S)$,
of $S$ as the {\it initial globally hyperbolic region}.) However, it is 
additionally required that there exists a compact set $K$ such that all
the  past-directed null generators of the future Cauchy horizon,
$H^+(S)$,  eventually enter and remain within $K$. It is not difficult
to see that any spacetime obtained by starting with a globally
hyperbolic spacetime and then smoothly deforming the metric in a compact
region so that the spacetime admits closed timelike curves must possess
a compactly generated Cauchy horizon. Conversely, as we  shall discuss
further in Sect.~\ref{geom}, any spacetime with a compactly  generated
Cauchy horizon necessarily violates strong causality, and, thus, is at
least `on the verge' of creating a time machine. Thus, we will, in the
following discussion,  identify the notions of `manufacturing a time
machine' and `producing a spacetime with a compactly generated Cauchy
horizon'.

It follows from the `Topological Censorship Theorem' 
\cite{FriedSchlWitt} that in order to produce a spacetime with a 
compactly generated Cauchy horizon by means of a traversable wormhole it
is necessary  to violate the weak energy condition. More generally, the
weak energy  condition must be violated in any spacetime with a 
compactly generated Cauchy horizon with noncompact partial Cauchy
surface $S$ \cite{Haw92}. Physically realistic classical matter fields
satisfy the weak energy condition, but this condition can be violated in
quantum field theory. Thus, `quantum  matter'  undoubtedly would be
needed to produce a spacetime with a compactly  generated Cauchy
horizon. 

The above considerations provide motivation for the study of  quantum
field theory on  spacetimes with a compactly generated Cauchy horizon.
In attempting to carry out such a study however, one immediately
encounters the problem that, even for linear field theories,
we only  have a clear and undisputed set of
rules for  quantum field theory in curved spacetime in the case that the
spacetime is globally hyperbolic.  In this case,
there is a well-established construction of a field algebra based on
standard theorems on the  well-posedness of the corresponding classical
Cauchy problem.  Friedman and Morris \cite{FriedMor} have recently
established that there exists a well defined classical dynamics on a 
simple model spacetime with closed timelike curves,  and they have
conjectured that classical dynamics will be suitably well posed on a
wide class of spacetimes with compactly generated Cauchy horizons.  Thus
for spacetimes  in this class one might expect it to be possible to
mimic  the standard construction and obtain a sensible quantum field
theory.

It should be noted that even if no difficulties  were to arise in the
formulation  of quantum field theory on spacetimes with compactly
generated Cauchy horizons, there still  would likely be very serious
obstacles  to manufacturing time machines, since it is far from clear
that any solutions to the semiclassical field equations can exist which
correspond to time machine production. In particular, not only the
(pointwise) weak  energy condition but the averaged null energy
condition must be violated with any time machine  produced with
`traversable wormholes' \cite{FriedSchlWitt}. Under some additional
assumptions, violation of the averaged null energy condition also must
occur in any  spacetime with a compactly generated Cauchy horizon in 
which the partial Cauchy surface $S$ is noncompact \cite{Haw92}.
Although the  averaged null energy condition  can be violated in quantum
field theory in curved spacetime,  there is recent evidence to suggest
that it may come `close enough' to holding  to provide a serious
impediment to the construction of a time machine
\cite{FordRoman,FlanWald}.

However, in the present paper, we shall not concern ourselves with 
issues such as whether sufficiently strong violations of energy 
conditions can occur to even create the conditions needed to produce a
spacetime with a compactly generated Cauchy horizon. Rather we will
focus on the more basic  issue of whether a sensible, nonsingular
quantum field theory of a linear field can be defined at all on such
spacetimes.  There is, of course, no difficulty in defining quantum
field theory in the initial globally hyperbolic region $D(S)$, but there
is evidence suggesting that quantum effects occurring as one approaches
the Cauchy horizon must become unboundedly large, resulting in singular
behaviour of the theory. Analyses by Kim and Thorne \cite{KimThorne} and
others \cite{Haw92,Visbk} have indicated that for all/many physically
relevant  states the renormalized expectation value of  the
stress-energy tensor,  $\left<T_{ab}\right>$, of a linear quantum field, defined on
the initial globally hyperbolic region $D(S)$, must blow up as one
approaches a  compactly generated Cauchy horizon.\footnote{\label{HK}
A first result in this direction was  obtained as early as 1982 by
Hiscock and Konkowski \cite{HiscKonk} who constructed a natural quantum 
state for a linear scalar field on the initial globally hyperbolic
portion of four dimensional Misner space  and showed that its stress
energy tensor diverges as one  approaches the Cauchy horizon.} However,
these arguments are heuristic in nature,  and examples recently
have been given by Krasnikov \cite{Krasnikov} and  Sushkov
\cite{Sushkov1,Sushkov2} of states for certain linear quantum field
models on the  initial globally hyperbolic region of (two and four
dimensional) Misner space  (see e.g., \cite{HawkEll} or \cite{Haw92}) for
 which $\left<T_{ab}\right>$ remains finite as one approaches the Cauchy horizon. 
This raises the issue of whether a quantum field necessarily becomes
singular at all on a compactly  generated Cauchy horizon, and, if so, in
what sense it must be `singular'.

The purpose of this paper is to give a mathematically  precise answer to
this question. As we shall describe further in  Sect.~\ref{geom},
every compactly  generated Cauchy horizon, $H^+(S)$, contains a nonempty
set, ${\cal B}$ of  `base points' having the property that every 
generator of $H^+(S)$ approaches arbitrarily close to ${\cal B}$ in the
past, and strong causality is violated at every $x \in {\cal B}$. We
shall prove the following two theorems concerning quantum field theory
on spacetimes with compactly generated  Cauchy horizons. (Full
statements are  given in Sect.~\ref{nogo}.   See also Sect.~\ref{disc} 
for further discussion.): 

\vspace{.5em}\noindent
{\bf Theorem 1:} {\it There is no extension to $(M,g_{ab})$ of the usual field 
algebra on the initial globally hyperbolic region  $ D(S)$ which
satisfies the condition of F-locality \cite{KayFLoc}.  (The F-locality
condition necessarily breaks down at any $x \in {\cal B}$.)}

\vspace{.5em}\noindent
{\bf Theorem 2:} {\it The two-point distribution for any Hadamard state of the
covariant Klein-Gordon field defined on the initial globally hyperbolic
region of a spacetime with a compactly generated Cauchy horizon must 
(when extended to a distributional bisolution on the full spacetime) be 
singular at every $x \in {\cal B}$ in the sense that the difference between 
this two point distribution and a local Hadamard distribution cannot be 
given by a bounded function in any neighbourhood  
(in $M \times M$) of $(x,x)$.} 
\vspace{.5em}

Theorem 1 shows that on a spacetime with a compactly generated Cauchy
horizon the quantum field theory must be locally different from the
corresponding theory on a globally hyperbolic spacetime.  Theorem 2
establishes that even if $\left<T_{ab}\right>$ remains finite as one approaches the
Cauchy horizon as in the examples of Krasnikov and Sushkov
\cite{Krasnikov,Sushkov1,Sushkov2}\footnote{\label{KrasSush}
Sushkov's examples involve a mild generalization of the quantum field
model discussed here to the case of  complex automorphic fields
\cite{Sushkov1,Sushkov2}.   While our theorems strictly don't apply as
stated to automorphic fields, we remark that we expect everything we do
to generalize to this case. (See also the final sentence of Sect.~2.)
Moreover, Cramer and Kay \cite{CramerKay}
have recently shown directly that the conclusions of Theorem 2 are valid
for the state discussed by Sushkov in \cite{Sushkov2}. Note that, as we
discuss further in Sect.~6 below, reference \cite{CramerKay} also points
out a similar situation for the states discussed by Boulware \cite{Boulware}
for Gott space and Tanaka and Hiscock \cite{TanakaHiscock} for Grant space.
}  it nevertheless
must always (i.e., for any Hadamard state on the initial globally
hyperbolic region) be the case that $\left<T_{ab}\right>$ is ill defined or
singular at all points of ${\cal B}$, since the limit which defines
$\left<T_{ab}\right>$ via the point-splitting prescription cannot exist,
and in fact must diverge in some directions.

Our theorems show that very serious  difficulties arise when attempting
to define the quantum field theory of a linear field on a spacetime with
a compactly generated Cauchy horizon. In  particular, our results may be
interpreted as indicating that in order to manufacture a time machine,
it would be necessary  at the very least to enter a regime where quantum
effects of gravity itself will be dominant. Thus, our results may be
viewed as supportive of Hawking's `Chronology Protection Conjecture'
\cite{Haw92}, although we shall refrain from speculating as to whether
the difficulties we find might somehow be evaded in a complete theory
where gravity itself is quantized.

The proofs of the above two theorems will be based upon the
`Propagation of Singularities' theorems of Duistermaat and H\"ormander
\cite{DuiHor,Horvol4}.  In particular, these theorems will allow us to
conclude that the two-point distribution of a Hadamard state is not
locally in  $L^2(M \times M)$ in a neighbourhood  of any $(x,y) \in M
\times M$ such that $x$ and $y$ can be joined by a null geodesic. This
fact directly gives rise to the singular behaviour of Theorem 2 at ${\cal
B}$. It should be noted that similar singular behaviour will occur in
essentially all (but -- see Sect.~\ref{disc} -- not quite all)
situations where one has a closed or `almost closed' or
self-intersecting null geodesic, so the results obtained here for 
spacetimes with compactly generated Cauchy horizons can be generalized 
to additional wide classes of causality violating spacetimes.   (See
Sect.~\ref{disc}.)

We shall begin in Sect.~\ref{geom} by establishing  two
geometrical propositions on spacetimes with compactly generated Cauchy
horizons.  In Sect.~\ref{micro}, after a brief review of the 
essential background on distributions and microlocal analysis  required
for their statement, we review the Propagation of Singularities theorems
of Duistermaat and H\"ormander for linear partial differential
operators. In Sect.~\ref{quantum} we shall briefly review 
some relevant aspects of linear quantum field theory in globally
hyperbolic spacetimes and discuss how one can  approach the question of
what it might mean to quantize a (linear) quantum field on a 
non-globally  hyperbolic spacetime. In particular, the notion of
F-locality, introduced in \cite{KayFLoc}, will be briefly reviewed
there.   In Sect.~\ref{nogo} we shall state and prove our theorems on
the singular behaviour of quantum fields.  In a final discussion section
(Sect.~\ref{disc}) we shall discuss further the significance and
interpretation of our theorems and mention some directions in which they
may be generalized.

We remark that while we explicitly treat only the covariant Klein-Gordon
equation, we expect appropriate analogues of Theorems 1 and 2 to hold
for arbitrary linear quantum field theories.   Moreover, it seems
possible that the singular behaviour we find for linear quantum fields 
on the Cauchy horizon may be related to pathologies found in the
calculation of the S-matrix for nonlinear fields \cite{FriedPapSim}.

\section{Some Properties of Compactly Generated Cauchy Horizons} \label{geom}

In this section, we shall establish some geometrical properties of  spacetimes
possessing compactly generated Cauchy horizons. We begin by recalling several
basic definitions and theorems, all of which can be found, e.g., in
\cite{HawkEll,WaldGen}.

Let $(M,g_{ab})$ be a time orientable spacetime, and let $S$ be a
closed, achronal subset of $M$. We define the {\it future domain of
dependence} of $S$ (denoted, $D^+(S)$) to consist of all $x \in M$
having the property that every past inextendible causal curve through
$x$ intersects $S$.  The past domain of dependence is defined similarly.
It  then follows that $\mbox{int\,} D^+(S) = I^-[D^+(S)] \cap I^+(S)$,
where $I^-$ denotes the chronological past. Thus, if $x \in D^+(S)$ but
$x \not\in S$, then every past directed timelike curve from $x$ 
immediately enters $\mbox{int\,} D^+(S)$ and  remains in $\mbox{int\,} D^+(S)$
until it intersects $S$.

The {\it future Cauchy horizon} of $S$ (denoted $H^+(S)$) is 
defined by $H^+(S) = \overline{D^+(S)} - I^-[D^+(S)]$, 
where $\overline{D^+(S)}$ denotes the closure of
${D^+(S)}$. It follows immediately that $H^+(S)$ is achronal and closed. A
standard theorem \cite{HawkEll,WaldGen} establishes that every $x \in H^+(S)$
lies on a null geodesic contained within $H^+(S)$ which either is past
inextendible or has an endpoint on the edge of $S$. Thus, if $S$ is edgeless
(in which case $S$ is referred to as a {\it partial Cauchy surface} or {\it
slice}), then $H^+(S)$ is generated by null geodesics which may have future
endpoints (i.e., they may `exit' $H^+(S)$ going into the future) but cannot
have past endpoints. Note that this implies that $H^+(S) \cap D^+(S)
= \emptyset$. Since similar results hold for $H^-(S)$, it follows that for
any partial Cauchy surface $S$, the full domain of dependence,
$D(S) \equiv D^+(S) \cup D^-(S)$, is open.

Now, let $S$ be a partial Cauchy surface. We say that  the future Cauchy
horizon, $H^+(S)$, of $S$ is {\it compactly generated} \cite{Haw92} if there
exists a compact subset, $K$, of $M$, such that for each past directed null
geodesic generator, $\lambda$, of $H^+(S)$ there exists a parameter value 
$s_0$ (in the domain of definition of $\lambda$) such that $\lambda (s) \in K$
for all $s > s_0$. In other words, when followed into the past, all null
generators of $H^+(S)$ enter and remain forever in $K$.

We now introduce some new terminology. Let $\lambda : I \rightarrow M$
be any continuous curve defined on an open interval, $I \subset \R$,  
which may be infinite or  semi-infinite.  We say that $x \in M$ is
a  {\it terminal accumulation point} of $\lambda$ if for every open
neighbourhood ${\cal O}$ of $x$ and every  $t_0 \in I$ there exists $t
\in I$ with $t > t_0$ such that  $\lambda (t) \in {\cal O}$.  (By
contrast, $x$ would be called an {\it endpoint} of $\lambda$  if for
every open  neighbourhood ${\cal O}$ of $x$ there exists $t_0 \in I$ 
such that $\lambda (t) \in {\cal O}$ for all $t > t_0$. Thus, an
endpoint is automatically a terminal accumulation point, but not
vice-versa.) Equivalently,  $x$ is a terminal accumulation point of
$\lambda$ if there exists a monotone increasing sequence  $\{ t_i \} \in
I$ without limit in $I$  such that $\lambda(t_i)$ converges to $x$. When
$\lambda$ is a causal curve, we shall call $x$ a {\it past terminal
accumulation point} if it is a terminal accumulation point when
$\lambda$ is parametrized so as to make it past-directed. Note that if
$x$ is a past terminal accumulation point of a causal curve $\lambda$
but $x$ is not a past endpoint of $\lambda$,  then strong causality must
be violated at $x$.

Let $H^+(S)$ be a compactly generated future  Cauchy horizon. We define
the {\it base set},  ${\cal B}$, of $H^+(S)$ by
\begin{eqnarray}
{\cal B} = \{x \in H^+(S) & | & \mbox{there exists a null generator,}\ \lambda,
\mbox{of } H^+(S) \ \mbox{such that}\  x \nonumber \\  
&& \mbox{is a past terminal accumulation point of}\ \lambda \}\;. \label{B} 
\end{eqnarray} 
Since the null generators of $H^+(S)$ cannot have past endpoints, it follows
that strong causality must be violated at each 
$x \in {\cal B}$. The following 
proposition (part of which  corresponds closely 
to the second theorem stated in
Sect.~V of \cite{Haw92}) justifies the terminology `base set':
  
\vspace{.5em}\noindent
{\bf Proposition 1.} {\it The base set, ${\cal B}$,  of any compactly
generated Cauchy horizon, $H^+(S)$, always is a nonempty subset of $K$.
In addition, all the generators of $H^+(S)$ asymptotically approach
${\cal B}$  in the sense that for each past-directed generator,
$\lambda$, of $H^+(S)$ and each open neighbourhood ${\cal O}$ of ${\cal
B}$, there exists a $t_0 \in I$ (where $I$ is the interval of definition
of $\lambda$) such that $\lambda (t) \in {\cal O}$ for all $t > t_0$.
Finally, ${\cal B}$ is comprised by null geodesic generators, $\gamma$,
of $H^+(S)$ which are contained entirely within ${\cal B}$ and  are both
past and future inextendible.}

\vspace{.5em}\noindent
{\it Proof.\/} Let $\lambda$ be a past directed null generator of
$H^+(S)$ and let $\{ t_i \}$ be any monotone increasing sequence without
limit on $I$.  Then $\lambda (t_i) \in K$ for infinitely many $i$. It
follows immediately that $\{\lambda (t_i) \}$ must have an accumulation
point,  $x$, which  must lie on $H^+(S)$ since $H^+(S)$ is closed.  It
follows that $x \in {\cal B}$, and hence ${\cal B}$ is nonempty. The
proof that ${\cal B}\subset K$ is entirely straightforward. If one could
find a generator $\lambda$ and an open neighbourhood ${\cal O}$ of ${\cal
B}$ such that for all $t_0 \in I$, we can find a $t > t_0$ such that
$\lambda (t) \not\in {\cal O}$, then, using compactness of $K$, we would
be able to find a  past terminal accumulation point of $\lambda$ lying
outside of ${\cal O}$ and, hence, outside of ${\cal B}$, which would
contradict the definition of ${\cal B}$.

To prove the last statement in the proposition, it is useful to introduce
(using the paracompactness of $M$) a smooth  Riemannian metric,
$e_{ab}$, on $M$. We shall parametrize all curves in $M$ by arc length,
$s$, with   respect to this Riemannian metric. 

Let $x \in {\cal B}$. We wish to show that there exists a null geodesic,
$\gamma$, through $x$ which is both past and future inextendible and is
contained in ${\cal B}$.  Since $x \in {\cal B}$, there exists a  null
generator, $\lambda$, of $H^+(S)$ such that $x$ is a past terminal
accumulation point of $\lambda$.  We parametrize $\lambda$ so that its
arc length parameter, $s$, increases in the past direction.  Since
$\lambda$ is past inextendible and $K$ is compact,
$s$ must extend to infinite values  (even
though the geodesic affine parameter of $\lambda$  in the Lorentz metric
$g_{ab}$ might only extend to finite values). Thus, there exists a
sequence $\{ s_i \}$ diverging to infinity such that $\lambda(s_i)$
converges to $x$.  Let $(k_i)^a$ denote the tangent to $\lambda$ at
$s_i$ in the arc length parametrization, so that $(k_i)^a$ has unit norm
with respect to $e_{ab}$. Since the subset of the tangent bundle of $M$
comprised by points in $K$ together  with tangent vectors with unit norm
with respect to $e_{ab}$ is compact,  it follows that -- passing to a
subsequence, if necessary -- there must exist a tangent vector $k^a$ at
$x$ such that $\{(\lambda(s_i), (k_i)^a) \}$ converges to $(x, k^a)$.
Since each $(k_i)^a$ is null in the Lorentz metric $g_{ab}$, it follows
by continuity that $k^a$ also must be null with respect to $g_{ab}$. 

Let $\gamma$ be the maximally extended (in $M$, in both future and 
past directions) null geodesic determined by $(x, k^a)$. We parametrize
$\gamma$ by arc length with respect to $e_{ab}$, with
increasing $s$ corresponding to going into the past and with 
$s = 0$ at $x$.   
Let $y \in \gamma$ and let
$s$ denote the arc length parameter of $y$. Since the sequence $\{ s_i
\}$ diverges to infinity, it follows that for sufficiently large $i$, $(s +
s_i)$ will be in the interval of definition of $\lambda$. Since
$\{(\lambda(s_i), (k_i)^a) \}$ converges to $(x, k^a)$, it follows by
continuity of both the 
exponential map (with respect to $g_{ab}$) and the arc length 
parametrization (with respect to $e_{ab}$) that $\{\lambda(s +s_i) \}$
converges to $y$. Thus, $y \in {\cal B}$, as we desired to show. $\Box$

In some simple examples (see, e.g., \cite{Visbk}), ${\cal B}$ consists of
a single closed null geodesic, referred to as a `fountain'. However, it
appears that, generically, ${\cal B}$ may contain null geodesics which
are not closed \cite{CI}.

Our final result on ${\cal B}$, which will be needed in Sect.~\ref{nogo}, is
the following. 

\vspace{.5em}\noindent
{\bf Proposition 2.} {\it Let $H^+(S)$ be a compactly generated Cauchy
horizon, and let ${\cal B}$ be its base set. Let $x \in {\cal B}$,  and let
${\cal U}$ be any globally hyperbolic 
open neighbourhood of $x$. Then there
exist points $y, z \in {\cal U} \cap \mbox{int\,} D^+(S)$ such that $y$ and $z$ 
are connected by a null geodesic in the spacetime $(M,g_{ab})$, but 
cannot be connected by a causal curve lying within ${\cal U}$.}

\vspace{.5em}\noindent
{\it Proof.\/} As in the proof of the previous proposition, 
we introduce  a smooth Riemannian metric, 
$e_{ab}$, on $M$ and parametrize curves in $M$ by arc
length, $s$, with  respect to $e_{ab}$ with
increasing $s$ corresponding to going into the past. 
Let $\lambda$ and  $\{ s_i \}$ be as in
the proof of the previous theorem. There cannot exist an $s_0$ such that
$\lambda(s) \in {\cal U}$ for all $s > s_0$, since otherwise strong causality
would be violated at $x$ in the spacetime $({\cal U}, g_{ab})$, thereby
contradicting the global hyperbolicity of $({\cal U}, g_{ab})$. Thus, there
exist integers $i, j$ with 
$s_i < s_j$ such that  $\lambda(s_i), \lambda(s_j) \in
{\cal U}$ but for some $s$ with $s_i < s < s_j$ we have $\lambda(s) \not\in
{\cal U}$. By passing to a smaller interval around
$s$ if necessary, we may assume
without loss of generality that $\lambda$ is one-to-one in $[s_i, s_j]$
(so that, in particular, $\lambda(s_i) \neq \lambda(s_j)$).
It then follows from the achronality of $H^+(S)$ that any past-directed
causal curve in $(M,g_{ab})$ which starts at $\lambda(s_i)$ and ends at
$\lambda(s_j)$ must coincide with (or contain)
this segment of $\lambda$. (If not, then there would exist two distinct,
past-directed, null geodesics connecting $\lambda(s_i)$ with
$\lambda(s_j)$, and it would be possible to obtain a timelike
curve joining $\lambda(s_i)$ to
$\lambda(s_j + 1)$.) Consequently, $\lambda(s_i)$ cannot be
joined to $\lambda(s_j)$ by any 
past-directed causal curve contained within ${\cal U}$.

Since in any globally hyperbolic spacetime, the causal past of any point is
closed, it follows that there exist open  neighbourhoods, ${\cal O}_i \subset
{\cal U}$ and ${\cal O}_j \subset {\cal U}$, of $\lambda(s_i)$ and of
$\lambda(s_j)$, respectively, such that no point of ${\cal O}_i$ can be 
joined to a point of ${\cal O}_j$ by a 
past-directed causal curve lying within ${\cal U}$. Without
loss of generality, we may assume  that ${\cal O}_i$ and ${\cal O}_j$ are
contained in $I^+(S)$ (since  
otherwise we could take their intersection with $I^+(S)$).

Our aim, now, is to deform $\lambda$ 
suitably to get a null geodesic in $(M,g_{ab})$ which joins a point 
$y \in {\cal O}_i \cap \mbox{int\,} D^+(S)$ to a
point $z \in {\cal O}_j \cap \mbox{int\,} D^+(S)$. To do so, we choose along
$\lambda$ (in a neighbourhood of $\lambda(s_i)$), a  
past-directed null vector
field, $l^a$, and a spacelike vector field $w^a$ satisfying $g_{ab} l^a k^b =
-1$, $g_{ab} w^a k^b = g_{ab} w^a l^b = 0$, and $g_{ab} w^a w^b = 1$, where
$k^a$ denotes the tangent to $\lambda$ 
(in the arc length parametrization with
respect to $e_{ab}$). Let $\epsilon > 0$ and consider the null geodesic
$\alpha$ starting at point  $\lambda(s_i + \epsilon)$ 
with null tangent  $k^a +\frac{1}{2} \epsilon^2 l^a + \epsilon w^a$. 
By continuity, for sufficiently
small $\epsilon$, $y = \alpha(\epsilon)$ will lie in ${\cal O}_i$ and $z =
\alpha(s_j - s_i)$ will lie in ${\cal O}_j$. 
Since $y$ and $z$ can be connected
to $\lambda(s_i)$ by a past-directed broken null geodesic, they each lie in
$I^-[H^+(S)]$. Since they also each lie in $I^+(S)$, it follows that $y,z \in
\mbox{int\,} D^+(S)$. As shown above, $y$
cannot be connected to $z$ by a past-directed causal curve lying within
${\cal U}$. However, since $y \in \mbox{int\,} D^+(S)$, it follows that
there cannot exist any future-directed causal curve,
$\sigma$, in $M$, connecting
$y$ to $z$, since, otherwise we would obtain a closed causal curve 
through $y$ by adjoining $\sigma$ to (the time reverse of) $\alpha$.
Thus, we have  $y,z \in {\cal U} \cap \mbox{int\,} D^+(S)$ such
that $y$ and $z$ are connected by the null geodesic $\alpha$ in the 
spacetime $(M,g_{ab})$, but they 
cannot be connected by any causal curve lying within
${\cal U}$. $\Box$

To end this section, we remark, following \cite{CramerKay}, that 
Propositions 1 and 2 (and hence also the theorems of Sect.~5)
will also clearly apply to any spacetime (such as four dimensional
Misner space) which, while having a Cauchy horizon which is not
compactly generated, arises as the product with a flat $4-d$ dimensional
Euclidean space of some spacetime with compactly generated Cauchy horizon
of lower dimension $d$.

\section{Microlocal Analysis and Propagation of Singularities} \label{micro}

In this section, we shall review results of 
H\"ormander and Duistermaat and H\"ormander 
\cite{DuiHor,Horvol3,Horvol4} 
concerning the propagation of singularities in
distributional solutions to partial differential equations. 
These results do not
appear to be widely known or used in the physics literature, and we shall
attempt to make them somewhat more accessible 
by stating them in the simpler
case of differential operators rather than in the more general setting of
pseudodifferential operators.

We begin by recalling that on an arbitrary smooth 
(paracompact) $n$-dim\-en\-sion\-al manifold, $M$,
elements of the vector space, ${\cal D}(M)$, 
of smooth ($C^\infty$) (and in this paper, we shall
assume real-valued) functions of compact support are referred to
as {\it test functions}.  A topology is defined on ${\cal D}(M)$ as follows. 

First, we introduce a Riemannian metric,  $e_{ab}$, and a derivative
operator, $\nabla_a$, on $M$. Next, we fix  a compact set,  $K$, and
focus attention on the subspace, ${\cal D}_K (M)$, of  ${\cal D}(M)$
consisting of test functions with support in $K$. On this subspace, we
define the family of  seminorms, $\{|| \cdot ||_k\}$, by  
\begin{equation}  ||f||_k = \sup_{x \in K}|\nabla_{a_1}...\nabla_{a_k}
f| \label{seminorm}\;,  \end{equation}  where by the `absolute value' of 
the tensor appearing on the right side of this equation we mean its norm
as computed using the Riemannian metric $e_{ab}$. We define the topology
of ${\cal D}_K (M)$ to be the weakest topology which make all of these
seminorms as well as the operations of addition and multiplication by
scalars continuous.  It may be verified that this topology is
independent of the  choices of $e_{ab}$ and $\nabla_a$. This gives each
${\cal D}_K (M)$  the structure of being a locally convex space.
Finally, we express $M$ as a countable union of compact sets, $K_i$
which form an increasing family ($K_i\subset K_{i+1}$) -- thereby
expressing ${\cal D} (M)$ as a countable union of the  ${\cal D}_{K_i}
(M)$ -- and take the topology of ${\cal D} (M)$ to be given by the
strict inductive limit \cite{ReeSiI} of the locally convex spaces ${\cal
D}_{K_i} (M)$.  It may be verifed that the topology thereby obtained on 
${\cal D} (M)$ is independent of the choice of compact sets $K_i$.

A distribution, $u$, on $M$, is simply a linear map from ${\cal D} (M)$ into
the real numbers, $\R$, which is continuous in the 
topology on ${\cal D}(M)$ defined in the  previous paragraph. 
The vector space of distributions on
$M$ is denoted ${\cal D}' (M)$.  Denote by 
$L^1_{\rm loc} (M)$ the collection
of measurable functions  on $M$ whose restriction 
to any compact set, $K$, is integrable with respect to a 
smooth volume element ${\bfeta}$ introduced on $M$.
(The definition of $L^1_{\rm loc} (M)$ clearly is independent 
of the choice of ${\bfeta}$.) If $F \in L^1_{\rm loc} (M)$, 
then the linear map $u: {\cal D}(M) \rightarrow \R$ given by 
\begin{equation} 
u(f) = \int_M F f {\bfeta}
\label{F} 
\end{equation} 
defines a distribution on $M$. We remark that in the
presence of a preferred volume element, 
such as provided by the natural volume element 
\begin{equation} 
{\bfeta}=|\mbox{det }g|^{1/2}dx^1 \wedge ... \wedge dx^n  \label{natvol}
\end{equation} 
associated with the metric $g_{ab}$ in the case where $M$ is a
spacetime manifold, we may identify 
the function $F$ with the distribution $u$.
A distribution $u \in {\cal D}' (M)$ will be said to be {\it smooth} if there
exists a $C^\infty$ function, $F$, on $M$ such that Eq.~(\ref{F}) holds. A
distribution, $u$, will be said to be {\it smooth at $x \in M$} if there exists
a test function, $g$, with $g(x) \neq 0$ such that $gu$ is smooth, where the
distribution $gu$ is defined by 
\begin{equation} 
gu(f) = u(gf)\;. 
\label{gu}
\end{equation} 
The set of points $y \in M$ at which $u$ fails to be smooth is
referred to as the {\it singular support} of $u$ in $M$.  

A key idea in the analysis of the propagation of 
singularities is to refine the notion of the singular support of $u$ in $M$ 
to a notion of the {\it wave front set} of $u$, which can be thought of 
as describing the singular support of $u$ in the {\it cotangent bundle}, 
$T^*(M)$, of $M$. This notion can be defined most
conveniently in terms of the Fourier transform of 
distributions. To begin, let
$u$ be a distribution on $\R^n$ which is of compact support, i.e., there
exists a compact set $K$ such that $u(f) = 0$ whenever 
the support of $f$ does not intersect $K$. We define the Fourier transform, 
$\hat{u}$ of $u$ to be the distribution given by 
\begin{equation} 
\hat{u}(f) = u(g \hat{f}) \label{ft}\;,
\end{equation} 
where $g$ is any test function such that $g = 1$ on $K$.
(The obvious extension of $u$ to act on complex 
test functions is understood here; $\hat{u}$ is a complex-valued  
distribution, but its real and
imaginary parts define real-valued  distributions.) It follows that $\hat{u}$
is always smooth (indeed, analytic), and the smooth function 
corresponding to $\hat{u}$ via Eq.~(\ref{F}) (which we also shall 
denote by $\hat{u}$) satisfies
the property that it and all of its derivatives are polynomially bounded at
infinity (see Theorem IX.5 of \cite{ReeSiII}). 
Furthermore, it follows from the
fact that the Fourier transform maps Schwartz space onto itself that the
distribution $u$ itself will be smooth if and only if for every  positive
integer $m$ there exists a constant $C_m$ such that 
\begin{equation} 
|\hat{u}(k)| \leq C_m (1 + |k|)^{-m}\;. 
\label{smooth1} 
\end{equation}

Now let $u \in {\cal D}' (M)$ be a distribution on an  
$n$-dimensional manifold, $M$. Let $x \in M$ and 
let ${\cal O}$ be an open neighbourhood of $x$ which can
be covered by a single coordinate patch, i.e., there exists a diffeomorphism 
$\psi : {\cal O} \rightarrow {\cal U} \subset \R^n$.  Let $g$ be a test
function with support contained within ${\cal O}$ such that 
$g(x) \neq 0$. The distribution $gu$ may then be 
viewed as a distribution on $\R^n$ which is
of compact support. Hence, for the given choice of coordinates, the Fourier
transform,  $\widehat{gu}$, of $gu$ is well defined as a 
distribution on $\R^n$ and satisfies the 
properties of the previous paragraph. We may use the
local coordinates at $x$ to identify the cotangent space at $x$ with $\R^n$ 
by associating with each cotangent vector $p_a$ the point in 
$\R^n$ given by the components of $p_a$ 
in these coordinates.  In this manner, we may
view $\widehat{gu}$ as a distribution on the cotangent space at $x$.

Now let $p_a$ be a nonzero cotangent vector at $x$. We say that $u$ is {\it
smooth} at $(x, p_a) \in T^*(M)$ if there exists a test function, $g$ with
support contained within ${\cal O}$ satisfying $g(x) \neq 0$ and 
there exists an open neighbourhood, ${\cal Q}$, 
of $p_a$ in the cotangent space at $x$ such
that for each positive integer $m$, there exists a constant 
$C_m$ such that for
all  $\rho_a \in {\cal Q}$ and all $\lambda \geq 0$ we have, 
\begin{equation}
|\widehat{gu}(\lambda \rho_a)| \leq C_m (1 + |\lambda|)^{-m}\;. 
\label{smooth2}
\end{equation} 
It can be shown that this notion of smoothness of $u$ at $(x,
p_a)$ is independent of the choice of local coordinates at $x$, and, thus, is a
well defined property of $u$ (see part (f) of Theorem IX.44 of 
\cite{ReeSiII}). 
Let ${\cal S} \subset T^*(M)$ denote the set of points in 
$T^*(M)$ at which $u$
is smooth. It follows directly  from its definition that ${\cal S}$ is open and
is `conic' in the sense that if $p_a \in {\cal S}$, then $\lambda p_a \in
{\cal S}$ for all  $\lambda > 0$. The {\it wave front set} of $u$, denoted
${\WF}(u)$, is defined to be the  complement of 
${\cal S}$ in $T^*(M) \setminus 0$, where `$0$' denotes the 
`zero section' of $T^*(M)$ 
\begin{equation}
{\WF}(u) = [T^*(M) \setminus 0] \setminus {\cal S} \:.
\label{WF} 
\end{equation} 
In other words, $(x, p_a) \in T^*(M)$ lies in ${\WF}(u)$ if and only if 
$p_a \neq 0$ and $u$ fails to be smooth 
at $(x, p_a)$. It can be shown (see, e.g., Theorem
IX.44 of \cite{ReeSiII}) that $x \in M$ is in the singular support of $u$ if
and only if there exists a cotangent vector $p_a$ at $x$ such that $(x, p_a)
\in {\WF}(u)$.

It is essential that we view ${\WF}(u)$ to be a subset of the cotangent
bundle (rather than, e.g., the tangent bundle) in order that it be 
independent of the choice of coordinates used to define Fourier
transforms. Some further insight into the meaning of the above definitions 
and  the reason why it is the cotangent bundle which is
relevant for the definition can be obtained from the
following considerations. Let $(x, p_a) \in T^*(M)$ with $p_a \neq 0$. Then
$p_a$ determines (by orthogonality) a  hyperplane in the 
tangent space at $x$.
In a sufficiently small open neighbourhood of $x$, we can introduce 
coordinates $(t, x^1,..., x^{n-1})$ so that the hypersurface of constant $t$
passing through $x$ is tangent to this hyperplane. 
These coordinates can be
used to factorize an open sub-neighbourhood of $x$ as the product manifold
$\R \times \R^{n-1}$.  Hence, any distribution, $u$,  defined on
this sub-neighbourhood can be viewed as a bi-distribution on 
$\R \times \R^{n-1}$.  In particular, for any test function, $f$, on 
$\R^{n-1}$, $u(\cdot, f)$ defines a distribution on $\R$. Then, it is not
difficult to verify that if a real-valued distribution $u$ is smooth at  $(x,
p_a) \in T^*(M)\setminus 0$, 
then there exists a $g\in{\cal D}(M)$ with $g(x) \ne 0$
such that  for any test function  $f$, on $\R^{n-1}$, the
distribution $gu(\cdot, f)$  on $\R$ is smooth. 

The notion of smoothness or the lack of smoothness of a 
distribution, $u$, at $(x, p_a) \in T^*(M)\setminus 0$ 
can be further refined as follows. First, for any real
number $s$, a distribution $u$ will be said to lie in the local Sobolev space
$H^s_{\rm loc}(x)$  associated with a point $x \in M$ if there exists  a test
function, $g$, with $g(x) \neq 0$ and the support of $g$ contained within a
single coordinate patch, such that $gu$  (viewed as a distribution 
of compact support on $\R^n$) satisfies 
\begin{equation} 
\int (1 + |k|^2)^s
|\widehat{gu}|^2 d^nk < \infty \;.
\label{sob} 
\end{equation} 
It is easy to see that the space so-defined is independent of the
choice of function $g$ and of the choice of coordinate patch  containing
its support. When $s$ is a non-negative  integer, this condition is
equivalent to requiring that $gu$ and all of its (weak)  derivatives up
to order $s$ are given by square integrable functions. (It is easy to
see that this result holds independently of the choice
of derivative operator and independently of the choice of (smooth)
volume element.)  Note that $u$ lies in $H^s_{\rm loc}(x)$ for all $s$
if and only if $u$ is smooth at $x$.  Following \cite{Horvol3}, we say
that a distribution $u$  lies in the local Sobolev space $H^s_{\rm
loc}(x, p_a)$ associated with  $(x, p_a) \in T^*(M)\setminus 0$ 
if we can express 
$u$ as $u = u_1 + u_2$, where $u_1$ lies in $H^s_{\rm loc}(x)$ and $(x,
p_a) \not\in {\WF}(u_2)$.  It can be shown (see Theorem 18.1.31 of
\cite{Horvol3}) that we have $u \in H^s_{\rm loc}(x)$ if and only if $u
\in H^s_{\rm loc}(x, p_a)$ for all nonvanishing cotangent vectors $p_a$
at $x$. Furthermore, we have $u \in H^s_{\rm loc}(x, p_a)$ for all  $s$
if and only if $u$ is smooth at $(x, p_a)$.

Next, we introduce some key definitions for linear partial differential 
operators. Let $A$ be an arbitrary linear partial differential operator of
order $m$ on $M$, so that $A$ can be expressed as 
\begin{equation} 
A =\sum_{i=0}^m \alpha^{a_1...a_i}_{(i)} \nabla_{a_1}...\nabla_{a_i} \;,
\label{pdo}
\end{equation} 
where $\nabla_a$ is an arbitrary derivative operator on $M$ and
each $\alpha^{a_1...a_i}_{(i)}$ is a smooth tensor field. We define the {\it
principal symbol}, $H$, of $A$ to be the map  
$H: T^*(M) \rightarrow \R$
given by 
\begin{equation} H(x, p_a) =\alpha^{a_1...a_m}_{(m)}(x)p_{a_1}...p_{a_m}\;.
\label{ps} 
\end{equation} 
It is easily checked that $H$ is independent of the choice of derivative
operator $\nabla_a$.

Now, $T^*(M)$ has the natural structure of a symplectic manifold, 
so it can be
viewed as the `phase space' of a classical dynamical system. By choosing
$H$ to be the Hamiltonian of this system, we thereby associate a classical
mechanics problem to each linear partial differential operator, 
$A$, on $M$. In
particular, associated with $A$, we obtain a vector field $h^a$ on $T^*(M)$
whose integral curves correspond to  solutions of Hamilton's equations of
motion on $T^*(M)$ with Hamiltonian $H$. We define the 
{\it characteristic set}
of $A$,  denoted ${\rm char}(A)$, to be the subset of $T^*(M) \setminus 0$
(where, again, $0$ denotes the zero-section of  
$T^*(M)$) satisfying $H(x, p_a)
= 0$. (In other words, the characteristic set of $A$ consists of the states in
phase space with zero energy but nonvanishing momentum.) 
We refer to the integral curves of 
$h^a$ in $T^*(M)\setminus 0$ starting from points in the characteristic
set as the {\it bicharacteristics} of $A$. (Often, these curves are called
`bicharacteristic strips', and the projection of a  bicharacteristic to $M$
is called a `bicharacteristic curve'.) By `conservation of energy', all
bicharacteristics are contained in the characteristic set.

We now are in a position to state the Propagation of Singularities Theorem
which will be used to prove our main results. First recall that, in the
presence of a preferred volume element ${\bfeta}$, 
if $A$ is a linear partial
differential operator, the adjoint of $A$, denoted $A^\dagger$, is
defined to be the linear partial differential operator
determined by the condition that
\begin{equation} 
\int g Af \bfeta = \int f A^\dagger g \bfeta 
\label{adjoint}
\end{equation} 
for all test functions $f,g$. 
A distribution $u$ will be said to satisfy the equation
$Au = 0$ if for every test function $f$, we have  
\begin{equation} 
u(A^\dagger f) = 0 \label{ua}\;.
\end{equation} 
We have the following theorem, which is obtained by restricting 
Theorems 26.1.1 and 26.1.4 of \cite{Horvol4} (from the case of
pseudodifferential operators) to
the simpler case of linear partial differential 
operators (and vanishing source term).

\vspace{.5em}\noindent
{\bf Propagation of Singularities Theorem.} {\it Let $M$ be an $n$-dimensional
manifold, with preferred volume element ${\bfeta}$.   Let $A$
be a linear partial differential operator of order $m$ on $M$ and
suppose  $u \in {\cal D}' (M)$ satisfies the equation $Au = 0$. Then, we
have (i) ${\WF}(u) \subset {\rm char}(A)$ and  (ii) For any  $(x, p_a)
\in {\rm char}(A)$,  we have $u \in H^s_{\rm loc}(x, p_a)$  if and only
if $u \in H^s_{\rm loc}(x', p'_a)$ for all $(x', p'_a)$ lying on the
same bicharacteristic as $(x, p_a)$. Thus, in particular,  if $(x, p_a)
\in {\WF}(u)$, then the entire bicharacteristic through $(x, p_a)$ lies
in ${\WF}(u)$.}
\vspace{.5em}

(Part (i) of the above theorem together with the final sentence
incorporate the content of Theorem 26.1.1 of \cite{Horvol4}, while Part
(ii) corresponds to Theorem 26.1.4.)

In the present paper, the partial differential operator in which we are
particularly interested is the covariant Klein-Gordon operator 
\begin{equation}
A=\Box_g - m^2 
\label{KG} 
\end{equation} 
on a given curved spacetime $(M,g_{ab})$.  Here, $\Box_g$ denotes the 
Laplace Beltrami operator for the
metric $g$.  We remark that if we take 
(as we shall from now on) our preferred
volume element ${\bfeta}$ to be the natural spacetime 
volume element (\ref{natvol})
associated with the metric, this operator satisfies 
$A^\dagger=A$. Clearly, its principal symbol is  
\begin{equation} 
H(x,p) = g^{ab}(x)p_ap_b
\end{equation} 
which is well known to be a Hamiltonian for geodesics. The
characteristic set thus consists of the points of $T^*(M)$ whose covector
is null and nonvanishing, and the bicharacteristics are curves $t\mapsto
(x(t), p_a(t))$ in the  cotangent bundle for which $t\mapsto x(t)$ is an
affinely parametrized null geodesic, and, at each value of the parameter
$t$, $p_a(t)$ is the cotangent vector obtained by using the metric to
`lower an index' on the tangent vector to the geodesic.  (Below, we
shall say that the cotangent vector $p_a(t)$ is `tangent' to the
geodesic.) In other words, in this case, the bicharacteristics are the
lifts to the cotangent bundle of affinely parametrized null geodesics.

In the proof of the theorems of the present paper we shall be  concerned
with certain distributional bisolutions to the covariant Klein-Gordon
equation which, as we shall discuss further in the next section,  occur
in quantum field theory on a curved spacetime $(M,g_{ab})$. The above
Propagation of Singularities Theorem will be used to obtain information
on the global nature of the singularities in these two-point functions
given information about the singularities when the two points on which
they depend are close together.   Roughly speaking, we shall be able to
conclude from the above theorem that, if such a distributional
bisolution is singular for sufficiently nearby pairs of points on a
given null geodesic, then it will necessarily remain singular for all
pairs of points on that null geodesic.  Moreover the theorem assures us
that the `strength' of the singularity (as measured by the indices of
the local Sobolev spaces in which the distribution fails to lie) cannot
diminish.  
  
One may define a distributional bisolution to be a  bidistribution on
$M$ which is a distributional solution to the covariant Klein-Gordon
equation in each variable.  Here, by a bidistribution $G$ on $M$ we mean
a (real or complex valued) functional on ${\cal D}(M)\times {\cal D}(M)$
which is separately linear and continuous in each variable and to say
that $G$ is a solution to the Klein-Gordon equation in each variable
means that $G((\Box_g -m^2)f,h)=0$ and $G(f,(\Box_g-m^2)h)=0$ for all
$f,h \in {\cal D}(M)$.    A bidistribution $G$ on $M$ is then
necessarily jointly continuous and arises from a distribution $\tilde G$
on the product manifold $M\times M$ in the sense that $G(f,g)=\tilde
G(f\otimes g)$, where, if $f$ and $g$ are each test functions in ${\cal
D}(M)$, $f\otimes g$ denotes the test function in ${\cal D}(M\times M)$ 
with values $f\otimes g(x,y)=f(x)g(y)$. (For the proof of these
assertions, see e.g., the proof of the Schwartz Kernel Theorem in Sect.~5.2 
in \cite{Horvol1}.) To say that $G$ is a distributional bisolution
on $M$ may thus be expressed by saying that $\tilde G$ is a
distributional solution to each of the pair of partial differential
equations $A_1\tilde G=0$, $A_2\tilde G=0$ on $M\times M$, where (in an
obvious notation) $A_1$ and $A_2$ are the partial differential operators
\begin{equation}  
A_1=(\Box_g-m^2)\otimes
I,\quad A_2=I\otimes (\Box_g-m^2)\;.
\end{equation}   
It is this latter way of regarding distributional bisolutions which permits
direct application of the Propagation of Singularity Theorem.  (From now on,
we shall adopt an informal point of view in which we do not distinguish between
$G$ and $\tilde G$.) Thus, a point $(x,p_a;y,q_b)$ in the cotangent bundle 
$T^*(M\times M)\setminus 0$ of $M\times M$ will belong to the characteristic set of both
$A_1$ and $A_2$ if and only if both $p_a$ and $q_b$ are null (and at least one
of them is non-zero).  Thus we conclude by Part (i) of the above theorem that
the wave front set of a distributional bisolution $\tilde G$ must consist of a
subset of such `doubly null' points. Moreover, if the wave front set of a
given distributional bisolution  $\tilde G$ includes such a doubly null point
$(x,p_a;y,q_b)$ then,  applying e.g., the last sentence of the Propagation
of Singularities Theorem to
the operator $A_1$ we  conclude that it must also include all points
$(x',p'_a;y,q_b)$ for which $(x',p'_a)$ lies on the same lifted null geodesic as
$(x,p_a)$. Similarly, applying the theorem to $A_2$, we conclude that it must
also include all points $(x,p_a;y',q'_b)$ for which $(y',q'_b)$ lies on the same
lifted null geodesic as $(y,q_b)$.
  
Next, we discuss some particular distributional bisolutions to the
covariant Klein-Gordon equation which will play an important role both
in our discussion of quantum field theory below, and in our theorems.
Firstly, given any globally hyperbolic spacetime $(M,g_{ab})$, then the
advanced and retarded fundamental solutions $\bigtriangleup_A(x,y)$, 
$\bigtriangleup_R(x,y)$ of the inhomogeneous Klein-Gordon equation exist
as 2-point distributions and are uniquely defined with respect to their
support properties \cite{Leray,Lichner,ChoquetBruhat}.  Their difference
$\bigtriangleup=\bigtriangleup_A-\bigtriangleup_R$ is then a preferred
distributional bisolution to the (homogeneous) covariant Klein-Gordon
equation.  It is clearly antisymmetric, i.e., 
$\bigtriangleup(f,g)=-\bigtriangleup(g,f)$ and, as we discuss in the
next section, plays the role of the `commutator function' in the
quantum theory.  One may show that its wave front set consists of {\it
all} elements $(x,p_a;y,q_b)$ of $T^*(M\times M) \setminus 0$ for which
$x$ and $y$ lie on a single null geodesic, for which $p_a$ and $q_b$ are
tangent to that geodesic, and for which $p_a$, when parallel transported
along that null geodesic from $x$ to $y$ equals $-q_a$.  (One way to
obtain this result is to notice that the advanced and retarded
fundamental solutions are special cases of advanced and retarded
`distinguished parametrices' in the sense of \cite{DuiHor} from which
the wave front set may be read off.  See also \cite{RadzPhD,RadzMicrolocal}.)

Secondly, for any curved spacetime $(M,g_{ab})$, we shall be interested in a
{\it class} of symmetric (i.e., $G(f,g)=G(g,f)$) distributional
bisolutions $G$ to the covariant Klein-Gordon equation which are  what
we shall call {\it locally weakly Hadamard}.  (These will arise in the
quantum field theory -- see Sect.~\ref{quantum} -- as (twice) the
symmetrized  two-point functions of `Hadamard states'.)  This notion
-- which is either weaker or equivalent to the various versions  of the
Hadamard condition which occur in the literature on quantum field theory
in curved spacetime -- is defined as follows: First, we require that $G$
be locally smooth for non-null related pairs of points in the sense that
every point $x$ in the spacetime  has a convex normal neighbourhood (see
e.g., \cite{HawkEll,WaldGen}) $N_x$ such that the  singular support of
$G$ in $N_x\times N_x$ consists only of pairs of null related points. 
In consequence of this (cf. the discussion around  Eq.~(\ref{F})
above) on the complement, $C_x$, in $N_x\times N_x$ of this singular
support (so $C_x$ consists of all pairs of non-null separated points in
$N_x\times N_x$) there will be a smooth two-point function, which we
shall denote by the symbol $G_s$, with the property that, for all test
functions $F$ supported in $C_x$,     
\begin{equation}
G(F)=\int_{C_x} G_s(y,z) F(y,z) {\bfeta}(y) {\bfeta}(z)  \;,
\end{equation} 
where we recall that ${\bfeta}$ denotes the natural volume
element (\ref{natvol}) associated with the metric.  It is easy to see
that, on $C_x$, $G_s$ must be a smooth bisolution to the covariant 
Klein-Gordon equation.  If $G$ is locally smooth for non-null related pairs of
points, then we say that $G$ is locally weakly Hadamard if for each
point $x$ in $M$, on the corresponding $C_x$, the $G_s$ as defined above
takes the `Hadamard form' \cite{Garab,DeWittBrehme,Wald77,Wald78,KayWald91}.
This latter condition is traditionally expressed by
demanding that there  exists some smooth function $w$ on $N_x \times
N_x$ such that the following equation holds on $C_x$  (in 4 dimensions,
with similar expressions for other dimensions)
\begin{equation} G_s(x,y)={1\over 2\pi^2}\left({\Delta^{1\over 2}\over\sigma} +
v\ln(|\sigma|)+w\right)  \;,
\label{Had} 
\end{equation} 
where $\sigma$ denotes the square of the geodesic distance between $x$ and $y$ 
(which is well defined in $C_x$ since $N_x$ is a convex normal neighbourhood),
$\Delta^{1\over 2}$ is the van Vleck-Morette determinant \cite{DeWittBrehme}
and $v$ is given by a power series in $\sigma$ with partial sums 
\begin{equation}  v^{(n)}(x,y)=\sum_{m=0}^n v_m(x,y)\sigma^m \;,
\end{equation}  where each $v_m$ is uniquely determined by the Hadamard
recursion relations  \cite{Garab,DeWittBrehme}.  This statement cannot be
interpreted literally because the power series defining $v$ does not in general
converge.  We overcome this problem and make the notion of {\it locally weakly
Hadamard} precise by replacing the above statement of Hadamard form by the
demand that, for each $x\in M$ and each integer $n$, there exists a $C^n$
function $w^{(n)}$ on $N_x \times N_x$ such that, on $C_x$   
\begin{equation}
G_s(x,y)={1\over 2\pi^2}\left({\Delta^{1\over 2}\over\sigma} +
v^{(n)}\ln(|\sigma|)+w^{(n)}\right)\;.  
\label{locweakHad} 
\end{equation}

The above notion of locally weakly Hadamard corresponds to the notion of
`Hadamard form' implicit in many references (e.g., 
\cite{Wald77,Wald78,Allen85}).  
However, we remark that the notion (sometimes referred
to as {\it globally Hadamard}) of `Hadamard' as defined (for the case
of globally hyperbolic spacetimes) in \cite{KayWald91} (when suitably
interpreted as a condition on a general symmetric distributional
bisolution on a globally hyperbolic spacetime) is a stronger notion
inasmuch as (a) it specifies, for each $x$, the local behaviour of the
distribution $G$ on test functions supported in $N_x \times N_x$ whose
support is not confined to $C_x$, by what amounts essentially to a
`principle part' prescription, (b) it explicitly rules out the
possible occurrence of so-called `non-local spacelike singularities'.
(See \cite{KayWald91} for details.)\footnote{\label{LocGlob} Actually, it
was conjectured by Kay \cite{KayComo,GonKay} and proved by Radzikowski
by microlocal analysis methods  \cite{RadzPhD,RadzLocGlob} that if the 
symmetrized two point function of a quantum state on the field algebra
(see Sect.~\ref{quantum}) for the covariant Klein-Gordon equation on a
globally hyperbolic spacetime satisfies the global Hadamard conditon of
Kay and Wald {\it locally} (i.e., on each element of an open cover) then
it is globally Hadamard. Thus, in the presence of the positivity
conditions required for a (symmetric) distributional bisolution (on a
globally hyperbolic spacetime) to be the symmetrized two-point function
of a quantum state, the strengthening of the Hadamard notion indicated
in point (b) here is automatic given that indicated in point (a).} 

Clearly, since (on any convex normal neighbourhood) $\sigma(x,y)$  is
smooth and vanishes if and only if $x$ and $y$ are null related,  the
singular support of any locally weakly Hadamard distributional
bisolution $G$, when restricted to $N_x \times N_x$ for any of the
neighbourhoods $N_x$ will consist precisely of all pairs of null related
points. Hence the wave front set of such a $G$, when restricted to
$T^*(N_x\times N_x) \setminus 0$ will include, for each such null pair,
at least one point $(y,p_a;z,q_b)$, where at least one of $p_a$ and $q_b$
is non-zero and where both $p_a$ and $q_b$ are null covectors which are
tangent to the null geodesic connecting $y$ and
$z$.\footnote{\label{symmetry} Of course, since $G$ is symmetric, if the 
point $(y,p_a;z,q_b)$ is in  its wave front set, then the point
$(z,q_b;y,p_a)$ will also be in its wave front set.} (If they were not
tangent, one could get a contradiction with the smoothness of $G$ at
non-null related pairs of points by  applying the Propagation of
Singularities Theorem.)\footnote{\label{WFSSC} More is known about the
wave front set of the (unsymmetrized) two-point functions of quantum
states (see Sect.~\ref{quantum}) on globally hyperbolic spacetimes
which are (globally) Hadamard in the
stronger sense of \cite{KayWald91}: Radzikowski \cite{RadzPhD,RadzMicrolocal} 
(see also
\cite{KohNewEx,BrunFredKoh,KohThesis,JunkThesis} for recent further
developments in this direction) has shown that the wave front set of any
such two point function consists precisely of all elements
$(x,p_a;y,q_b)$ of $T^*(M\times M) \setminus 0$ for which $x$ and $y$
lie on a single null geodesic, for which $p_a$ is tangent to that null
geodesic and future pointing, and for which $q_a$, when parallel
transported along that null geodesic from $y$ to $x$ equals $-p_a$.}   

In fact, one can show more than this:  Namely, given any (symmetric)
locally weakly Hadamard distributional bisolution $G$, for each point
$x$ in $M$ and each pair of null related points $(y,z)\in N_x\times N_x$
 with $y \neq z$ there exist null covectors $p_a$ at $y$ and $q_b$ at
$z$ which are each tangent to the null geodesic connecting $y$ and $z$
and which are not both zero (see Footnote \ref{symmetry}) such that $G$
fails to belong to  $L^2_{\rm loc}(y,p_a;z,q_b)$ (i.e., to $H^0_{\rm
loc}(y,p_a;z,q_b)$).  To prove this result, it suffices to show that
$1/\sigma$, and hence, easily, also $G$ fails to  belong to $L^2_{\rm
loc}(y,z)$ for any pair, $(y,z)$, of distinct null-related points in
$N_x\times N_x$, since then we have 
$G \not\in L^2_{\rm loc}(y,p_a;z,q_b)$ for some  
covectors $p_a$ at $y$ and $q_b$ at $z$ which are restricted by 
arguments of the previous paragraph. However, the fact that
$G \not\in L^2_{\rm loc}(y,z)$ follows 
immediately from the following lemma [where we
identify $L$ with $M\times M$, $h$ with $\sigma$, and $X$ with $(y,z)$]:

\vspace{.5em}\noindent
{\bf Lemma.} {\it Let $L$ be a manifold, and let $h$ be a smooth function on 
$L$ which vanishes
at a point $X \in L$ but whose gradient is nonvanishing at $X$. 
Then $1/h$ (or, more precisely, any distribution which
agrees with $1/h$ for $h \neq 0$) fails to be locally $L^2$ at $X$.}
\vspace{.5em}

(The proof is immediate once one chooses a coordinate 
chart around $X$ in which
the function $h$ is one of the coordinate functions.)

\section{The Quantum Covariant Klein-Gordon Equation on a Curved Spacetime}
\label{quantum}

In our discussion of quantum field theory, 
we shall restrict our interest to a
linear Hermitian scalar field, satisfying the covariant Klein-Gordon equation 
\begin{equation} 
(\Box_g - m^2)\phi=0 
\label{KGeq} 
\end{equation} 
on a curved spacetime $(M,g_{ab})$.  We shall now outline a 
suitable mathematical description of this theory 
in terms of the algebraic approach to quantum field
theory.  For further discussions of this, and closely related, approaches see
e.g., \cite{KayWald91,KayFLoc,WaldQuant}.

In the  case that $(M,g_{ab})$ is globally hyperbolic, we may take the field algebra
to be the $*$-algebra with identity $I$ generated by polynomials in `smeared
fields' $\phi(f)$, where $f$ ranges over the space $C_0^\infty(M)$ of
smooth real valued functions compactly supported on $M$, which satisfy the
following relations (for all $f_1, f_2 \in C_0^\infty(M)$ and for all pairs of
real numbers $\lambda_1,\lambda_2$): 
\begin{enumerate} 
\item $\phi(f)=\phi(f)^*$

\item $\phi(\lambda_1f_1+\lambda_2f_2)=\lambda_1\phi(f_1)+\lambda_2
\phi(f_2)$

\item $\phi((\Box_g-m^2)f)=0$

\item $[\phi(f_1),\phi(f_2)]=\I\bigtriangleup(f_1,f_2)I$, \end{enumerate}
where $\bigtriangleup$ denotes the classical `advanced minus retarded'
fundamental solution (or `commutator function')  discussed in the
previous section.  (Of Relations (1)-(4), it is thus only Relation (4)
which becomes problematic when we attempt to go beyond the class of
globally hyperbolic spacetimes.  We shall return to this point below in
our discussion of `F-locality'.)  

To be precise, what we mean by the above statement is that we regard the
set of polynomials, over the field of complex numbers,  of the abstract
objects, $\phi(f)$, with $f\in C_0^\infty(M)$ as a free $*$-algebra with
identity and then  quotient by the $*$-ideal generated by the above
relations. 

We have referred above to $\phi(f)$ as a `smeared quantum field'.  While,
in our mathematical definition above, this is to be thought of as a single
abstract object, we of course interpret it heuristically as related to the
`field at a point' `$\phi(x)$' by 
\begin{equation}
\phi(f)=\int_M\phi(x)f(x) {\bfeta} \;,
\end{equation} 
where ${\bfeta}$ is the natural volume element 
(\ref{natvol}) defined in the
previous section.  Of course we proceed in this way since the `field at a
point' is not expected by itself to be a mathematically well defined entity. 
This failure of the `field at a point' to exist is of course closely related
to the singular nature of the commutator function discussed 
in the previous section.

Quantum {\it states} are defined to be positive, normalized (i.e.,
$\omega(I)=1$) linear
functionals on this field algebra.  (Here, a state $\omega$ is said to be
positive if we have $\omega(A^*A)\ge 0$ for all $A$ belonging to the field
algebra.)  A state $\omega$ is thus specified by specifying the set of all its
`smeared $n$-point functions'  
\begin{equation}
\omega(\phi(f_1)\dots\phi(f_n)) \;.
\end{equation} 
One expects states of interest to at least be sufficiently regular for these
smeared $n$-point functions to be distributions -- i.e., one expects the
expression above to be a continuous functional of each of the quantities
$f_1,\dots, f_n$ when the space $C_0^\infty(M)$ is topologized in the ${\cal D}$
topology. By the Schwartz Kernel Theorem (see 
the previous section) we may
equivalently regard the $n$ point functions as distributions on 
$M\times \cdots \times M$.

By condition (4) above, for any state $\omega$, twice the antisymmetric
part of the two point function, $\omega(\phi(f_1)\phi(f_2))$, is simply 
$\I \bigtriangleup(f_1,f_2)$, which is smooth at all $(y,z)$ which cannot
be connected by a null geodesic.  Furthermore, for the reasons
discussed, e.g., in \cite{KayWald91,KayFLoc,WaldQuant} and briefly
reviewed below,  we require that (twice) the symmetrized two-point
function (i.e., $\omega(\phi(f_1)\phi(f_2)+\phi(f_2)\phi(f_1))$   should
(at least) be a locally weakly Hadamard distributional bisolution as
defined in Sect.~\ref{micro}. In this paper, we shall refer to states
satisfying this condition as `Hadamard states'.  Note that this notion
only restricts the two-point function and does not restrict the other
$n$-point functions; we shall not need to concern ourselves here with
the question of what  should be required of the short distance behaviour
of other $n$-point  functions in order for a state to be physically
realistic.

The main reason for requiring that a state satisfy this 
Hadamard condition is that it is necessary in order that
the following `point-splitting procedure' yield well-defined, finite
values at each point $y$ for
quantities such as the renormalized expectation value in that
state of $\phi^2$ or of the quantum stress-energy tensor $T_{ab}$:\footnote{
Note also that, on a globally hyperbolic spacetime, the quasi-free Hadamard
states satisfy a number of desirable properties \cite{Ver} including
local quasiequivalence.}
We define the expectation value of $\phi^2(y)$ at a point $y$ by
taking the neighbourhood $N_y$ of $y$ as in the previous section and setting
\begin{equation} 
\omega(\phi^2(y))=\lim_{(x,x')\rightarrow (y,y)}{1\over
2}(\omega(\phi(x)\phi(x')+\phi(x')\phi(x))-H^{(n)}(x,x'))\;,
\label{split} 
\end{equation} 
where $H^{(n)}(x,x')$ is an appropriate locally constructed Hadamard
parametrix, i.e., it is a function -- defined on the neighbourhood
$C_y$ consisting of all pairs of non-null related pairs of points in
$N_y$ -- of the form (\ref{locweakHad}) with a particular, locally defined
algorithm used to obtain $w^{(n)}$ (see, e.g., \cite{WaldQuant} for further
discussion).  In Eq.~(\ref{split}), it is understood that, before
taking the limit, each of the terms in the outer parentheses is
defined initially on $C_y$ (where they make sense as smooth
functions).  Because the state is assumed to satisfy the above
Hadamard condition, the full term in parentheses will then clearly
extend to a continuous (in fact, $C^n$) function on $N_y\times N_y$,
thus ensuring that the limit will be well defined.  There is a
similar, but more complicated, formula corresponding to (\ref{split})
for $\omega(T_{ab}(y))$ involving suitable (first and second)
derivatives with respect to $x$ and $x'$ in the terms in parentheses
and also involving the addition of a certain {\it local correction
term} \cite{Wald78,WaldQuant}.  The resulting prescription for
$\omega(T_{ab}(y))$ then satisfies a list of desired properties which
uniquely determines it up to certain finite renormalization
ambiguities \cite{Wald77,WaldQuant}.  This justifies the use of the
point-splitting procedure, thus leading to the conclusion that the
(locally weakly) Hadamard condition on a state must be satisfied in
order to ensure that the expectation values of the stress-energy
tensor be well-defined.\footnote{\label{weakerHad} In fact, since one
only requires the difference between the two point function and the
locally constructed Hadamard parametrix, $H$, to be $C^2$, it would
clearly suffice for the well-definedness of expectation values of the
renormalized stress-energy tensor to replace the condition of being
locally weakly Hadamard (see before (\ref{locweakHad})) by a weaker
condition where one only demands that $w^{(n)}$ be $C^2$ for $n >
2$. We remark that it is easy to see that our Theorem 2 would continue
to hold with such a further weakening of the Hadamard condition.}

Next we turn to consider what it might mean to quantize the covariant
Klein-Gordon equation on a spacetime $(M,g_{ab})$ which is {\it not}
globally hyperbolic. Our approach will be to postulate what might be
regarded as `reasonable candidates for minimal necessary conditions'
for any such theory.  In other words, we consider statements which begin
with the phrase `Whatever else a quantum field theory (on a given 
non-globally hyperbolic spacetime) consists of, it should at least involve
$\dots$'  We shall consider independently two candidate conditions
of this type:

\vspace{.5em}\noindent 
{\bf Candidate Condition 1.} Whatever else a quantum field theory
consists of, it should at least involve a field algebra satisfying
{\it F-locality}  \cite{KayFLoc}.  In other words (see \cite{KayFLoc} for more
details) it should involve a field algebra which is a star algebra
consisting of polynomials in `smeared quantum fields' $\phi(f)$
which, just as in the globally hyperbolic case, satisfies the Relations
(1), (2) and (3) listed above.  Additionally, (this is the F-locality
condition of \cite{KayFLoc}) one postulates that Relation (4) (which, as
we mentioned above is the one relation for which the assumption of
global hyperbolicity is needed) should still hold in the following local
sense: {\it Every point in $M$ should have a globally hyperbolic
neighbourhood ${\cal U}$ such that, for all  $f_1, f_2\in
C_0^\infty({\cal U})$, Relation (4) holds with $\bigtriangleup$ replaced
by $\bigtriangleup_{\cal U}$}, where, by $\bigtriangleup_{\cal U}$, we
mean the advanced minus retarded fundamental solution for the region
${\cal U}$,  regarded as a globally hyperbolic spacetime in its own
right.\footnote{\label{Flocalquibbles}  In \cite{KayFLoc} the F-locality
condition was stated slightly differently; namely  that every
neighbourhood of every point in $M$ should contain a globally hyperbolic
subneighbourhood ${\cal U}$ such that, for all  $f_1, f_2\in
C_0^\infty({\cal U})$, Relation (4) holds with $\bigtriangleup$ replaced
by $\bigtriangleup_{\cal U}$.  However, it is clear from the F-locality
in this latter sense of the usual field algebra on a globally hyperbolic
spacetime (see \cite{KayFLoc}) that this is equivalent to the condition
given here.}
\vspace{.5em} 
 
In defence of such a condition, let us simply say here (see
\cite{KayFLoc} for further discussion) that it is motivated by the
philosophical bias (related to the equivalence principle) that, on an
arbitrary spacetime, the `laws in the small' for quantum field theory
should be the same as the familiar laws for globally hyperbolic
spacetimes. We remark that it is easy to see that the familiar laws for
globally hyperbolic spacetimes, as given above, are themselves F-local. 
It is also known that there do exist some non-globally hyperbolic
spacetimes which admit field algebras satisfying  F-locality.  (In the
language of \cite{KayFLoc}, there exist some non-globally hyperbolic
F-quantum compatible spacetimes.) In
particular there do exist  F-quantum compatible spacetimes with closed timelike
curves, for example the {\it spacelike cylinder} -- i.e., the region of
Minkowski space  (say with Minkowski coordinates $(t,x,y,z)$) between 
two times -- say $t_1$ and $t_2$ -- with the $(x,y,z)$ coordinates of
opposite edges identified.   (See \cite{KayFLoc} for the case of the
massless Klein-Gordon equation, and  \cite{FewHig} for the massive
case.)

Of course, as anticipated in \cite{KayFLoc}, the above philosophical bias
could be used to argue for a slightly different and possibly weaker
locality notion. In this connection, we remark that, since a first
version of this paper was written, evidence has emerged \cite{Fewsteretal}
that the examples of F-quantum compatible chronology violating spacetimes
mentioned above are unstable in the sense that there are arbitrarily
small perturbations of these spacetimes which fail to be F-quantum
compatible.  On the other hand, as is also pointed out in
\cite{Fewsteretal}, if one replaces the condition of F-locality by the
weaker notion of F-locality modulo $C^\infty$\footnote{I.e., for which the
above italicized definition holds when one replaces $\Delta_{\cal U}$ by 
$\Delta_{\cal U}+F$ for some (antisymmetric) $F\in C^\infty({\cal
U}\times{\cal U})$.}, then these examples of allowed chronology violating
spacetimes would become stable and there would be many other stable
examples (i.e., of chronology violating spacetimes which admit field
algebras which are F-local modulo $C^\infty$).

\vspace{.5em}\noindent 
{\bf Candidate Condition 2.} Whatever else a quantum field  theory
consists of, it should at least involve a field algebra satisfying
Relations (1), (2) and (3) listed above, and, in addition, there should
exist states for which (twice) the  symmetrized two-point function is a
locally weakly Hadamard distributional bisolution $G(f_1, f_2)$.
\vspace{.5em}  
 
The motivation for requiring the existence of a field algebra satisying
(1)-(3) is, of course, the same as for Candidate Condition 1. One also
can motivate the requirement of the existence of Hadamard states by a
philosophical bias similar to that motivating the F-locality condition:
The theory should admit `physically acceptable states', and these
physically acceptable states should have the same local character as in
the globally hyperbolic case.  However, there is additional strong
motivation for requiring the existence of Hadamard states: Let $(M,
g_{ab})$ and $(M', {g'}_{ab})$ be spacetimes -- one or both of which may
be non-globally-hyperbolic -- for which there exist open regions ${\cal
O} \subset M$ and ${\cal O}' \subset M'$ which are isometric. Let
$\omega$ be a state on the field algebra of $(M,g_{ab})$ and let
$\omega'$ be a state on the field algebra of $(M', {g'}_{ab})$. Use the
isometry between ${\cal O}$ and ${\cal O}'$ to identify these two
regions. Then, it is natural to postulate that -- under this
identification -- for all $y \in {\cal O}$ the {\it difference} between
$\omega(T_{ab}(y))$ and  $\omega'(T_{ab}(y))$ should be given by the
point-splitting formula in terms of the difference between the
symmetrized two-point functions of $\omega$ and $\omega'$ whenever this
formula makes sense; furthermore, when this formula does not make sense,
the difference between $\omega(T_{ab}(y))$ and  $\omega'(T_{ab}(y))$ is
ill defined or singular. (This  postulate may be viewed as a
generalization to non-globally-hyperbolic spacetimes of the main content
of the stress-energy axioms (1) and (2) of \cite{WaldQuant}.) If so, and
if in the globally hyperbolic case $\omega(T_{ab}(y))$ is given by the
point-splitting prescription as described above, then the locally weakly
Hadamard condition on the symmetric part of the two-point function -- or
slight weakenings thereof (see Footnote \ref{weakerHad}) -- must be
satisfied in order to have an everywhere defined, nonsingular
$\omega(T_{ab})$. If  $\omega(T_{ab})$ were singular for {\it every}
state $\omega$, the theory clearly would be unacceptable on physical
grounds, since singular `back reaction' effects would necessarily occur
for all states,  thereby invalidating the original background spacetime
upon which the quantum field  theory was based.

\section{Theorems} \label{nogo}

We are now ready to state and prove our main theorems, which  establish
that neither of the two Candidate Conditions of the previous section can
be satisfied by a Klein-Gordon field on a spacetime, $(M,g_{ab})$, with
compactly generated Cauchy horizon. These theorems are direct 
consequences of the Propagation of
Singularities Theorem of Sect.~\ref{micro} (applied to the
relevant distributional bisolutions) in combination with the geometrical
property of base points expressed in Proposition 2 of Sect.~\ref{geom}.  

\vspace{.5em}\noindent
{\bf Theorem 1.} {\it There is no extension to $(M,g_{ab})$ of the usual
field  algebra on the initial globally hyperbolic region  $D(S)$ which
satisfies F-locality at any base point (see Sect.~\ref{geom}) of the
Cauchy horizon.}

\vspace{.5em}\noindent
{\it Proof.\/} Let $x \in {\cal B}$ and let ${\cal U}$ be any globally 
hyperbolic neighbourhood of $x$. To prove the claimed violation of 
F-locality, it clearly suffices to prove that the restrictions of 
$\bigtriangleup_{D(S)}$ and $\bigtriangleup_{\cal U}$ to ${\cal U}\cap 
D(S)$ cannot coincide, where $\bigtriangleup_{D(S)}$ denotes the
advanced minus retarded fundamental solution for the initial globally
hyperbolic region  $ D(S)$, and $\bigtriangleup_{\cal U}$ denotes the 
advanced minus retarded fundamental solution for ${\cal U}$. However,
this follows immediately  from the fact that these two quantities cannot
take the same values at the pair of points $(y,z)$ of Proposition 2 of
Sect.~\ref{geom}.  Indeed, since $(y,z)$ are spacelike related in the
intrinsic geometry of ${\cal U}$, $\bigtriangleup_{\cal U}$ must clearly
vanish at this pair of points.  On the other hand, it follows from the
explicit description of the wave-front set of the advanced minus
retarded fundamental solution on any globally hyperbolic spacetime, as
given in Sect.~3, that $\bigtriangleup_{D(S)}$ must be singular at the
pair $(y,z)$ because they  are null related in the spacetime $ D(S)$.
$\Box$

By repeating this argument with $\bigtriangleup_{D(S)}(F_1,F_2)$
replaced by the commutator
$\omega(\phi(F_1)\phi(F_2)-\phi(F_2)\phi(F_1))$ and using
the Propagation of Singularities Theorem, the following closely
related theorem also can be readily proven:

\vspace{.5em}\noindent
{\bf Theorem 1$'$.} {\it Under the mild extra technical condition that the algebra
admits at least one state $\omega$ for which the smeared two point function
$\omega(\phi(f_1)\phi(f_2))$ is distributional, then there is no field
algebra whatsoever which satisfies F-locality (i..e., in the language of
\cite{KayFLoc}, spacetimes with compactly generated Cauchy horizons are {\it
non-F-quantum compatible}.)}
\vspace{.5em}

We remark that it is easy to see that Theorems 1 and 1$'$ will continue to
hold if one replaces the notion of F-locality by the weaker notion (see
Sect.~4) of F-locality modulo $C^\infty$.  (See \cite{Fewsteretal} for
further discussion.)

We note that, in the very special case of the massless two-dimensional
Klein-Gordon equation on two-dimensional Misner space,  Theorems 1 and
1$'$ had been obtained previously by relying on the explicitly known
propagation of the two-dimensional wave equation.  (See \cite{KayFLoc}.)

The following theorem establishes that Candidate Condition 2
cannot hold:

\vspace{.5em}\noindent
{\bf Theorem 2.} {\it Let $G$ be a distributional bisolution on $(M,g_{ab})$
which is everywhere  locally weakly Hadamard on the initial globally
hyperbolic region $ D(S)$. Then $G$ fails to be locally weakly Hadamard
at any $x \in {\cal B}$ in the following severe sense: The difference
between $G$ and any  locally constructed Hadamard parametrix $H^{(n)}$
(see Eq.~(\ref{split}) above) will fail to be given by a locally
$L^2$  function on $N\times N$, where $N$ is any convex normal
neighbourhood of $x$ (so that $H^{(n)}$ is well defined  on $N\times N$).
Thus, for any $n$, $G - H^{(n)}$ cannot be given by a continuous,  nor
even by a bounded function on $N\times N$, and quantities such as the
renormalized expectation value of $\phi^2$ or the renormalized
stress-energy tensor must be singular or ill defined at any base point
of the Cauchy horizon.}

\vspace{.5em}\noindent
{\it Proof.\/} Let ${\cal U}$ be a globally hyperbolic subset of $N$
containing $x$. Let $(y,z)$ be as in Proposition 2 of Sect.~\ref{geom}, 
so that $y$ and $z$ are spacelike-separated in $({\cal U},
g_{ab})$ but are joined by a null  geodesic, $\alpha$, in $M$.  Let $y'
\in {\cal U}$ lie along  $\alpha$ sufficiently near to $y$ to be
contained within the convex normal neighbourhood of $y$ appearing in the
locally weakly Hadamard property of $G$. Then we know by the discussion
at the end of Sect.~\ref{micro} that  $G$ must fail to belong to the
space $L^2_{\rm loc}(y,p_a;y',p'_b)$ for some pair $p_a$ at $y$ and
$p'_b$ at $y'$ of null tangents to this null geodesic, which moreover
cannot both be zero.   Assuming that $p_a$ does not vanish (otherwise,
in what follows replace ($y,p_a$) by ($y',p'_a$)) we may conclude by
Part (ii) of the Propagation of Singularities Theorem of Sect.~\ref{micro} 
that $G$ cannot belong to the space $L^2_{\rm
loc}(y,p_a;z,q_b)$ for some null covector $q_b$ at $z$. It follows that
$G$ cannot belong to $L^2_{\rm loc}(y,z)$.  On the other hand, $H^{(n)}$ is
non-singular at the pair $(y,z)$, since $y$ and $z$ are
spacelike-separated in ${\cal U}$.   We conclude that $G-H^{(n)}$ cannot
arise from a locally $L^2$ function on $N\times N$. $\Box$

It is clear from the proof of Theorem 2 that the failure to be locally
$L^2$ must already occur if one restricts attention to the part of the
neighbourhood $N$ which lies in the initial globally hyperbolic region
$D(S)$. In fact we have:

\vspace{.5em}\noindent
{\bf Theorem 2$'$ (slightly stronger than Theorem 2).} 
{\it Let $G$ be a distributional
bisolution on $(M,g_{ab})$ which is everywhere  locally weakly Hadamard
on the initial globally hyperbolic region $D(S)$ and let $N$ be any
convex normal neighbourhood of any $x\in {\cal B}$.  
Then $G-H^{(n)}$ fails to be given
by a  locally $L^2$ function on $N\cap D(S)$.}
\vspace{.5em}

This is now a statement entirely about the behaviour of the quantum theory
on the initial globally hyperbolic region $D(S)$ as one approaches a base
point. 
Thus it seems fair to conclude from this theorem that something must go
seriously wrong with the quantum field theory on the Cauchy horizon
almost independently of any assumptions (cf. e.g., `Candidate Condition
2' in the previous section) about what would constitute an extension of
the quantum field theory beyond the initial globally hyperbolic region.

It is also worth remarking that nowhere in the proof of Theorem 2 have
we made any  use of the positivity conditions (see e.g.,
\cite{KayComo,GonKay,KayFLoc}) required of a symmetric distributional
bisolution (on a globally hyperbolic spacetime) in order for it to arise
as (twice) the symmetrized two-point function of a quantum state on the
field algebra of Sect.~\ref{quantum}. Also, it should be noted that 
Theorem 2 would continue to hold if one were to weaken the notion of
`locally weakly Hadamard' along the lines indicated in  Footnote
\ref{weakerHad}. 

Finally we recall (see end of Sect.~2) that the theorems of this section
will also hold for any spacetime (such as four dimensional Misner space)
which arises as the product with a Euclidean $4-d$ space of a spacetime
with compactly generated Cauchy horizon of lower dimension $d$.

\section{Discussion} \label{disc}

In this section, we make some remarks concerning the significance of our
theorems, and also discuss a number of directions in which these
theorems may be further generalized.

First, we attempt to clarify the significance of our theorems by
contrasting the situation for quantum field theory on spacetimes with
compactly generated Cauchy horizons with that on spacetimes with Killing
horizons.  Theorems 1 and 2 here are concerned with a particular
instance of a situation where one asks about the extension of a quantum
field theory (in our case, the covariant Klein-Gordon field) from some
given globally hyperbolic spacetime $(N,h_{ab})$ to a larger spacetime
$(M,g_{ab})$.   In the situation addressed by Theorems 1 and 2, $(M,g_{ab})$ is a
spacetime with compactly generated Cauchy horizon, and  $(N,h_{ab})$ its
initial globally hyperbolic region $D(S)$.   However, another, familiar
(see e.g., \cite{Fulling,Unruh,BirrellDavies,KayWald91,WaldQuant})
instance of such a  situation  is the case where $(M,g_{ab})$ is Minkowski
spacetime  and $(N,h_{ab})$ the Rindler wedge. (Equally, we could take the
case where $(M,g_{ab})$ is the Kruskal spacetime and $(N,h_{ab})$ the exterior
Schwarzschild spacetime, etc.)  It is interesting to contrast these two
situations.  In the Minkowski-Rindler wedge case, of course, both
spacetimes are globally hyperbolic and it is clear that the field
algebra one would construct for the Rindler wedge -- regarding it as a
globally hyperbolic spacetime in its own right -- is naturally
identified with the subalgebra of the Minkowski spacetime field algebra
associated with the  Rindler wedge region. Thus the field algebra on
Minkowski spacetime (which is, of course, F-local) certainly constitutes
an F-local extension of the field algebra for the Rindler wedge.  Thus
there is no analogue of Theorem 1 in this case.  Let us next turn to the
question of whether there are any Hadamard states on the Rindler wedge
algebra whose symmetrized two point distributions extend to everywhere
locally weakly Hadamard (symmetric) distributional bisolutions  on the
whole of Minkowski space.  (We shall refer to this, from now on, as the
question whether `Hadamard states have Hadamard extensions').  It is
certainly the case that {\it most} Hadamard states on the Rindler wedge
have {\it no}  Hadamard extension to the whole of Minkowski space.  For
example, the  Fulling vacuum (i.e., the ground state for the
one-parameter group of wedge-preserving Lorentz boosts), as well as the
KMS states with respect to the group of wedge-preserving Lorentz boosts
at all `temperatures' except $T=1/2\pi$ (which corresponds to the
Minkowski vacuum state) fail to have Hadamard
extensions.\footnote{\label{KWtheorem} In fact, as proven in
\cite{KayWald91,KaySuff}, the Minkowski vacuum state is the only
globally Hadamard state (satisfying a `no zero mode' condition)
on Minkowski space which is globally invariant
under the same group of Lorentz boosts and an analogous result holds for
the analogous Kruskal-Schwarzschild situation and a wide range of other
analogous situations involving spacetimes with bifurcate Killing
horizons.} In fact, it is well known that, for all these states, the
expectation value of the stress-energy tensor diverges as one approaches
the horizon.  Nevertheless, there do of course exist Hadamard  states on
the Rindler wedge which have  Hadamard extensions to the whole of
Minkowski spacetime and for which the stress-energy tensor is bounded: 
namely the restrictions to  the Rindler wedge algebra of Hadamard states
on Minkowski space!  Prior to the results in the present paper, it was
unclear to what extent the situation was analogous for Hadamard
extensions of Hadamard states on the initial globally hyperbolic region
$ D(S)$  of a spacetime $(M,g_{ab})$ with a compactly generated Cauchy
horizon. The work of Kim and Thorne \cite{KimThorne}, Hawking
\cite{Haw92}, Visser \cite{Visbk}, and others strongly suggested that
(just as in the Rindler-Minkowski situation) {\it most} Hadamard states
on $ D(S)$  have a stress-energy tensor which diverges on the Cauchy
horizon.   However, there were also examples -- albeit in the context of
two-dimensional models \cite{Krasnikov},  or for special models
involving automorphic fields  \cite{Sushkov1,Sushkov2} 
(see Sect.~\ref{intro}) -- of states on $ D(S)$ 
for which the stress-energy tensor
vanished.   Thus one might have thought that (as in the
Rindler-Minkowski situation) there could still be  some/many Hadamard
states on $ D(S)$ with Hadamard extensions to  $(M,g_{ab})$. Theorem 2 proves
that this is not the case.  Furthermore, while it  does not rule out the
possible existence of Hadamard states on $D(S)$  for which the stress
energy tensor is bounded, it still implies that any such state must have
a stress-energy tensor which is singular {\it at} the base points of the
Cauchy horizon.  In this important sense, the situation for spacetimes
with compactly generated Cauchy horizons is thus quite distinct from the
Minkowski-Rindler situation.

Finally, we point out that theorems similar to Theorems 1 and 2 will 
clearly hold in any spacetime (not necessarily with a compactly generated 
Cauchy horizon) which contains an almost closed (but not closed) null
geodesic or a self-intersecting (but not closed) null geodesic (since,
clearly, the same conclusions that hold for the base points of Proposition
2 of Sect.~2 will hold for any accumulation point, respectively for any 
intersection point). Thus, in particular, analogues of 
Theorems 1 and 2 will hold for points on the `polarized hypersurfaces' in 
the time-machine models discussed by Kim and Thorne \cite{KimThorne}, Gott
\cite{Gott}, Grant \cite{Grant}, and others since these contain 
self-intersecting null geodesics.  For further discussion, see
\cite{CramerKay} where it is pointed out that, because of the accumulation
of polarized hypersurfaces at the chronology horizons of these models,
analogues of Theorem 2 -- but not Theorem 2$'$ 
-- will hold for points on the chronology horizons of Gott and
Grant space. As remarked in \cite{CramerKay}, this result holds
notwithstanding the existence of the states
exhibited by Boulware \cite{Boulware} (for sufficiently
massive fields on the initial
globally hyperbolic region of Gott space) and Tanaka and Hiscock
\cite{TanakaHiscock} (for sufficiently
massive fields on the initial globally
hyperbolic region of Grant space) for which the stress-energy tensor is
bounded (i.e., on the initial globally hyperbolic region).  Moreover
analogues of Theorems 1 and 2 are expected to hold in {\it most\/}
spacetimes which contain a closed null geodesic since generically, the
same conclusions as hold for the base points in Proposition 2 of Sect.~2
will hold for each point on such a geodesic. However, we remark
that there {\it do} exist very special cases of spacetimes with closed
null geodesics for which (for suitable field theories) F-local field
algebras do exist and everywhere locally Hadamard (symmetric)
distributional bisolutions {\it do} exist.  One such special spacetime
is the double covering of compactified Minkowski space.  (We are
grateful to Roger Penrose for pointing this example out to us.) The
special feature of this spacetime which makes it possible to
evade the conclusions of Proposition 2 of Sect.~2 and hence to
evade the arguments of Theorems 1 and 2 is that the entire light cone through
any point, when globally extended, refocusses back onto that point. It
is not difficult to see that, on this spacetime, the field algebra
obtained by conformally mapping the field algebra for the massless Klein
Gordon equation in Minkowski space extends to an F-local field algebra
(i.e., for the conformally coupled massless Klein-Gordon equation).
Equally, (twice) the conformally mapped symmetrized two-point function
for the massless Klein-Gordon equation on Minkowski space extends, on
this spacetime to an everywhere locally Hadamard distributional
bisolution (again of the conformally coupled massless Klein-Gordon
equation).   A two-dimensional example with similar behaviour is
provided by the two-dimensional massless Klein-Gordon equation on the
two-dimensional `null strip' -- i.e., the region between two parallel
null lines in two-dimensional Minkowski space with opposite edges
identified (by identifying points intersected  by the same null lines).

\section{Acknowledgements}
We wish to thank Piotr Chrusciel for bringing
Theorem 26.1.4 of \cite{Horvol4} to our attention. We also thank Roger Penrose
for pointing out the example mentioned in Sect.~6. B.S.K. thanks Stephen
Hawking for conversations in which he raised some of the issues addressed
in this work.
M.J.R. wishes to  thank the University of Toronto for hospitality at a
late stage of this work.
This work was supported by NSF grants PHY-92-20644 
and PHY-95-14726 to the University 
of Chicago and by EPSRC grant GR/K 29937 to the University of York.

\noindent
Communicated by A.~Connes 
\end{document}